\documentclass[pra,floatfix,showpacs,amsmath,twocolumn]{revtex4}
\usepackage{amssymb}
\usepackage{graphics}
\usepackage{epsfig}
\usepackage{psfrag}
\begin{document}

\newcommand{\be}{\begin{eqnarray}}
\newcommand{\ee}{\end{eqnarray}}
\newcommand{\bea}{\begin{eqnarray}}
\newcommand{\eea}{\end{eqnarray}}
\newcommand{\bma}{\begin{subequations}}
\newcommand{\ema}{\end{subequations}}
\def\qed{\leavevmode\unskip\penalty9999 \hbox{}\nobreak\hfill
     \quad\hbox{\leavevmode  \hbox to.77778em{%
               \hfil\vrule   \vbox to.675em%
               {\hrule width.6em\vfil\hrule}\vrule\hfil}}
     \par\vskip3pt}
\def\lR{l^2_{\mathbb{R}}}
\def\RR{\mathbb{R}}
\def\E{\mathbf e}
\def\D{\boldsymbol \delta}
\def\S{{\cal S}}
\def\T{{\cal T}}
\def\dd{\delta}
\def\id{{\bf 1}}
\newcommand{\Eg}{E_G}
\newcommand{\tr}[1]{{\rm tr}\left[#1\right]}
\newtheorem{theorem}{Theorem}
\newtheorem{prop}{Prop.}
\newtheorem{definition}{Def.}
\newtheorem{lemma}{Lemma}
\newcommand{\C}{{\Bbb C}}
\newtheorem{cor}{Corollary}
\title{Gaussian Entanglement of Formation}

\author{M. M. Wolf$^{1,2}$, G. Giedke$^{1,3}$,
O. Kr\"uger$^{2}$, R. F. Werner$^{2}$, and J. I. Cirac$^{1}$}
\affiliation{\footnotesize (1) Max-Planck-Institut f\"ur
Quantenoptik,
Hans-Kopfermann-Str. 1, Garching, D-85748, Germany\\
(2) Institut f{\"u}r Mathematische Physik, Mendelssohnstr.~3,
D-38106
Braunschweig, Germany\\
(3) Institut f{\"u}r Quantenelektronik, ETH Z{\"u}rich,
Wolfgang-Pauli-Stra{\ss}e 16, CH-8093 Z{\"u}rich, Switzerland}

\date{\today}

\begin{abstract}
We introduce a Gaussian version of the entanglement of formation
adapted to bipartite Gaussian states by considering decompositions
into pure Gaussian states only. We show that this quantity is an
entanglement monotone under Gaussian operations and provide a
simplified computation for states of arbitrary many modes. For the
case of one mode per site the remaining variational problem can be
solved analytically. If the considered state is in addition
symmetric with respect to interchanging the two modes, we prove
additivity of the considered entanglement measure.
Moreover, in this case and considering only a
single copy, our entanglement measure coincides with the true
entanglement of formation.
\end{abstract}

\maketitle

\section{Introduction}
One of the main novelties of {\it Quantum Information Theory} is
to consider entanglement no longer merely as an apparent
paradoxical feature of correlated quantum systems, but rather as a
resource for quantum information processing purposes. This new
point of view naturally raises the question regarding the
quantification of this resource. How much entanglement is
contained in a given state? For pure bipartite states there is,
under reasonable assumptions, a simple and unique answer to this
question, namely the von Neumann entropy of the reduced state
\cite{DHR02,Vid99,PR97}. For mixed states there are several
entanglement measures \cite{Hor01}, which can be distinguished due
to their operational meaning and mathematical properties. Such a
measure should be non-increasing under mixing as well as under
local operations and classical communication (LOCC), and it should
return the right value for pure states. The largest measure
fulfilling these requirements is the {\it Entanglement of
Formation} $E_F$ \cite{BDSW96}. Operationally, it quantifies the
minimal amount of entanglement, which is needed in order to
prepare the state by mixing pure entangled states. It is therefore
defined as an infimum
\begin{equation}\label{EoFdef}
E_F(\rho)=\inf \left\{\sum_k p_k E(\Psi_k)\Big|\; \rho=\sum_k p_k
|\Psi_k\rangle\langle\Psi_k|\right\}
\end{equation} over all (possibly continuous) convex decompositions  of the
state into pure states with respective entanglement $E(\Psi)=
S\big({\rm tr}_B[|\Psi\rangle\langle\Psi|]\big)$, where
$S(X)=-\tr{X\log X}$ is the von Neumann entropy.  By its
definition calculating $E_F$ is a  highly non-trivial optimization
problem, which becomes numerically intractable very rapidly if we
increase the dimensions of the Hilbert spaces. Remarkably, there
exist analytical expressions for two-qubit systems \cite{Woo98} as
well as for highly symmetric states \cite{TV00,VW00}.

Recently, $E_F$ was calculated for the first time for continuous
variable states, namely for symmetric Gaussian states of two modes
\cite{GWKWC03}. In general, Gaussian states are distinguished
among other continuous variable states due to several reasons.
Experimentally, they are relatively easy to create and arise
naturally as states of the light field of a laser
(cf.\cite{VWW01}) or in atomic ensembles interacting with light
\cite{JKP00}. For this and other reasons they play a more and more
important role in Quantum Information Theory (cf. \cite{BP02}).

Theoretically, despite the underlying infinite dimensional Hilbert
space, they are completely characterized by finitely many
quantities -- the first and second moments of canonical operators.
Moreover, they stand out due to several extremal properties
\footnote{E.g., Gaussian states have
  maximal entropy among all states with given first and second moments
  of the canonical observables.}. In fact, the calculation of $E_F$
for symmetric two-mode Gaussian states depends crucially on the
fact that for given ``EPR-correlations'' two-mode squeezed
Gaussian states are the cheapest regarding entanglement. This
implies that in this particular case there is a decomposition in
terms of pure Gaussian states, which is optimal for $E_F$ in
(\ref{EoFdef}).

On the one hand, this raises the question, whether this is
generally true for all Gaussian states. On the other hand one may,
motivated by the operational interpretation of $E_F$, restrict
Eq.~(\ref{EoFdef}) to decompositions into Gaussian states from the
very beginning. After all, Gaussian states arise naturally,
whereas the experimental difficulties of preparing an arbitrary
pure continuous variable state are by no means simply
characterized by the amount of its entanglement. For these reasons
we will in the following investigate the {\it
  Gaussian Entanglement of Formation} $E_G$ to quantify the
entanglement of bipartite Gaussian states by taking the infimum in
Eq.~(\ref{EoFdef}) only over decompositions into pure Gaussian states.

This article is organized as follows: in Sec.~\ref{statesSec} we
recall basic notions concerning Gaussian states. Sec.~\ref{EGSec}
defines the {\it Gaussian Entanglement of Formation} and provides
a major simplification concerning its evaluation for bipartite
Gaussian states of arbitrary many modes. In Sec.~\ref{monSec} we
prove that $E_G$ is indeed a (Gaussian) entanglement monotone, in
the sense that it is non-increasing under Gaussian local
operations and classical communication (GLOCC). The case of
general two-mode Gaussian states is solved analytically in
Sec.~\ref{1x1genSec}. The special case of symmetric Gaussian
states, for which it was proven in \cite{GWKWC03} that $E_G=E_F$
is discussed in detail in Sec.~\ref{symSec}, where we give an
alternative calculation of $E_G$, which is in turn utilized in
Sec.~\ref{addSec} in order to prove additivity of $E_G$ for this
particular case. Finally, Sec.~\ref{expSec} applies the measure to
some examples which arise when a two-mode squeezed state is sent
through optical fibers. The appendix proves a Lemma about
decompositions of classical Gaussian probability distributions.

\section{Gaussian states}\label{statesSec}
Consider a bosonic system of $n$ modes, where each mode is
characterized by a pair $Q_k, P_k$ of canonical (position and
momentum) operators. If we set $R=(Q_1, P_1,\ldots,Q_n,P_n)$ the
canonical commutation relations are  governed by the {\it
symplectic
matrix}\begin{equation}\label{sigma} \sigma\equiv\bigoplus_{k=1}^n \left(%
\begin{array}{cc}
  0 & 1 \\
  -1 & 0 \\
\end{array}%
\right)
\end{equation} via $[R_k,R_l]=i\sigma_{kl}$. A state is called a Gaussian state
if it is completely characterized by the first and second moments
of the canonical operators $R_k$ in the sense that the
corresponding Wigner function is a Gaussian. Utilizing Weyl
displacement operators $W_{\xi}\equiv e^{i\xi^T\sigma R}$, the
first moments $d_k\equiv\tr{\rho R_k}$ can be changed arbitrarily
by local unitaries. Hence, all the information about the
entanglement of the state is contained in the covariance matrix
(CM)
\begin{equation}\label{covdef}
\gamma_{kl}\equiv\tr{\rho \{R_k-d_k\id, R_l - d_l\id \}_+},
\end{equation}where $\{\cdot,\cdot \}_+$ denotes the
anti-commutator. By definition the matrix $\gamma$ is real and
symmetric, and due to Heisenberg's uncertainty relation it has to
satisfy $\gamma\geq i\sigma$. For pure Gaussian states we have
$\det(\gamma)=1$ or, equivalently, $(\sigma\gamma)^2=-\id$.

In the following we  denote  the density operator
corresponding to the Gaussian state with covariance matrix
$\gamma$ and displacement vector $d$ by $\rho_{(\gamma,d)}$. If
the latter is a bipartite state, its tensor product structure
corresponds to a partition of the $n$ modes into two subsets.

An important decomposition of $\rho_{(\gamma,0)}$ into pure
Gaussian states is given by
\begin{equation}\label{Gaussiandecomp}
\rho_{(\gamma,0)}\propto\int d^{2n}\xi\  \rho_{(\gamma_p,d-\xi)}\
e^{-\frac14 \xi^T(\gamma-\gamma_p)^{-1}\xi}
\end{equation}
where $\gamma_p\leq\gamma$ is the covariance matrix
of a pure Gaussian state. Since displacements of the form
$\rho_{(\gamma,0)}\mapsto \rho_{(\gamma,d)}$ are local operations,
Eq.~(\ref{Gaussiandecomp}) tells us that starting with
$\rho_{(\gamma',0)}$ we can obtain every Gaussian state with CM
$\gamma\geq\gamma'$ by means of LOCC operations.

\section{Gaussian entanglement of formation}\label{EGSec}

We define the {\it Gaussian Entanglement of Formation} $E_G$ for a
bipartite Gaussian state $\rho_{(\gamma,d)}$ by
\begin{eqnarray}\label{EGdef}
E_{G}\big(\rho_{(\gamma,d)}\big)&\equiv&
\inf_{\mu}\bigg\{\int\mu(d\gamma_p,dD)\; E(\rho_{(\gamma_p,D)})\ \Big|  \\
&&\quad \rho_{(\gamma,d)}=\int\mu(d\gamma_p,dD)
\rho_{(\gamma_p,D)} \bigg\}\label{EGdecomp} ,
\end{eqnarray}
where the infimum is taken over all probability measures $\mu$
characterizing convex decompositions of $\rho_{(\gamma,d)}$ into
pure Gaussian states $\rho_{(\gamma_p,D)}$, and
$E(\rho_{(\gamma_p,D)})$ is the von Neumann entropy of the reduced
state. For pure $n\times n$ mode Gaussian states this quantity can
be readily expressed in terms of the symplectic eigenvalues of the
reduced CM. Denote these eigenvalues by $a_k, k=1,\dots,n$. We
have $a_k\geq1$ and define $r_k\geq 0$ by $a_k=\cosh r_k$. Then
\begin{equation}\label{tmss_ent}
E(\rho_{(\gamma,d)}) = \sum_k H(r_k),
\end{equation}
where
\begin{equation}\label{Hr}
H(r) = \cosh^2(r)\log_2(\cosh^2r)-\sinh^2(r)\log_2(\sinh^2r).
\end{equation}
To obtain this expression note that any pure $n\times n$ Gaussian
state is locally equivalent (via unitary GLOCC)  to the tensor product
of $n$ two-mode squeezed states with squeezing parameters $r_k$
\cite{HW01}. For each tensor factor the entanglement is given by the
above formula.  Since the symplectic spectrum of the reduced CM is
invariant under local unitary GLOCC the $r_k$ can be computed
directly from $\gamma_p$ as described above.

The integrals in Eqs.~(\ref{EGdef}, \ref{EGdecomp}) are taken over
the space $\mathbb{R}^n$ of displacements and over the set of
admissible pure state covariance matrices.
The following proposition tells us that it is sufficient to only
consider measures $\mu$, which vanish for all but one covariance
matrix:

\begin{prop}\label{propgen} The Gaussian entanglement of formation for a
bipartite Gaussian state $\rho_{(\gamma,d)}$ is given by
\begin{equation}\label{theoEoFG}
E_G\big(\rho_{(\gamma,d)}\big)=\inf_{\gamma_p} \Big\{
E(\rho_{(\gamma_p,0)})\ \big|\ \gamma_p\leq \gamma\Big\},
\end{equation}where the infimum is taken over pure Gaussian states
with CM $\gamma_p$.
\end{prop}
{\it Proof: } The proof can be divided into three steps:\\

(i) The problem can be reformulated in terms of classical Gaussian
distributions, by considering  Wigner functions instead of density
operators. This is formally achieved by taking the trace of the
decomposition in Eq.~(\ref{EGdecomp}) with the phase space
displaced parity operator ${\cal P}_{\xi}\equiv W_{\xi}{\cal P}
W_{\xi}^*$ \cite{Gro76}. Then
\begin{equation}\label{Wigner}
\tr{{\cal
P}_{\xi}\rho_{(\gamma,d)}}=|\gamma|^{-\frac12}\exp{\Big[-\frac14
(\sigma\xi+d)^T \gamma^{-1}(\sigma\xi+d)\Big]}
\end{equation} is up to a normalization factor equal to the Wigner function of
$\rho_{(\gamma,d)}$, which in turn completely determines the
state.

(ii) All the states $\rho_{(\gamma_p,D)}$ contributing to
Eqs.~(\ref{EGdef}, \ref{EGdecomp}) must have smaller covariance
matrices $\gamma_p\leq\gamma$, i.e.
\begin{equation}\label{muzero}
\mu\bigg(\Big\{(\gamma_p,D)\ \big|\ \exists_x : \langle x,
(\gamma-\gamma_p)x\rangle<0\Big\}\bigg)=0.
\end{equation}
The idea of the proof is that the tails of a Gaussian with CM that
is too large would give rise to an increasing and in the end
overflowing contribution if we only move far enough away from the
center. This is mathematically formalized in Lemma
\ref{classGauss} stated and proven in the appendix. To apply Lemma
\ref{classGauss} via Eq.~(\ref{Wigner}) to Eq.~(\ref{EGdef}) we need
in addition, that the inverse is operator monotone on positive
matrices $\gamma>0$.

(iii) Assume that $\tilde{\mu}$ is a measure corresponding to an
optimal decomposition of $\rho_{(\gamma,d)}$ giving rise to the
infimum in Eq.~(\ref{EGdef}). Then
\begin{eqnarray}
E_G (\rho_{(\gamma,d)}) &=&
\int_{\gamma_p\leq\gamma}\tilde{\mu}(d\gamma_p,dD) E(\rho_{(\gamma_p,D)})\\
&\geq& \inf_{\gamma_p}\Big\{ E(\rho_{(\gamma_p,0)})\ \big|\
\gamma_p\leq \gamma\Big\}\label{erg}.
\end{eqnarray}
However, by using a Gaussian decomposition of the form in
Eq.~(\ref{Gaussiandecomp}) we know that equality
 in Eq.~(\ref{erg}) can be achieved for a measure $\tilde{\mu}$ which is
Gaussian in $D$ and a delta function with respect to $\gamma_p$.
\qed

Proposition \ref{propgen} considerably simplifies the calculation
of $E_G$, since the optimization is reduced from the set of all
possible decompositions to the set of pure states satisfying the
matrix inequality $\gamma_p\leq\gamma$. Before we proceed to
calculate $E_G$ analytically for the two-mode case, we will
show that $E_G$ is indeed a proper entanglement monotone. To confine
the argument to CMs (rather than density matrices) we make use of
Proposition \ref{propgen}.

\section{Monotonicity under Gaussian operations}\label{monSec}
For $E_G$ to serve as a good Gaussian entanglement measure it
should not increase under GLOCC. That this is the case is quickly
seen using the characterization of Gaussian operations given in
\cite{GC02,Fiu02b}. There it was shown that the change of the CM
$\gamma$ of a Gaussian state under Gaussian operations takes the
form of a Schur complement:
\begin{equation}
G(\gamma) =
\tilde\Gamma_1-\tilde\Gamma_{12}\frac{1}{\tilde\Gamma_2+\gamma}
\tilde\Gamma_{12}^T\equiv S(\tilde\Gamma+0\oplus\gamma).
\end{equation}
Here
\[
\Gamma = \left( \begin{array}{cc}
\Gamma_1   & \Gamma_{12}\\
\Gamma_{12}^T &  \Gamma_2
\end{array} \right)
\]
is the $4n\times4n$ CM of the state characterizing the operation $G$ and
$\tilde\Gamma\geq0$ is the CM of the partially transposed state
\footnote{Partial transposition of an $n\times n$ Gaussian density
  matrix with CM $\gamma$ yields a Gaussian operator with CM
  $\tilde\gamma\equiv \Lambda\gamma\Lambda$, where
  $\Lambda=\id_{2n}\oplus\mathrm{diag}(1,-1,1,\dots,1,-1)$.}.

For positive matrices $A\geq B$ implies $S(A)\geq S(B)$ (cf.
\cite[p.472]{HJ87}). Consequently, if $\gamma \geq \gamma_p$, then
the transformed CM fulfills $G(\gamma)\geq G(\gamma_p)$.

Every GLOCC can be decomposed into a {\it pure} GLOCC $G_p$
mapping pure states onto pure states, and the addition of
classical Gaussian noise. This decomposition can easily be shown
using the above mentioned ordering of the Schur complements for
ordered matrices. The decomposition then reads
$G(\gamma_p)=G_p(\gamma_p)+P$, where the noise is characterized by
some positive matrix $P\geq 0$, which is usually state-dependent.
Therefore we have $G(\gamma)\geq G(\gamma_p)\geq G_p(\gamma_p)$
and since the latter CM corresponds to a pure state, which can be
obtained from $\gamma_p$ by a local Gaussian operation, its
entanglement is certainly smaller than that of $\gamma_p$
\cite{GECP03}. It follows that $E_G$ cannot increase under GLOCC.

\section{The general two mode case}\label{1x1genSec}
Now that we have assured that $E_G$ is a good measure of
entanglement in a Gaussian setting, we set to compute it for the
case of two Gaussian modes in an arbitrary mixed state. The CM
$\gamma$ of any two-mode Gaussian state can be brought to the
normal form \cite{DGCZ99,Sim00}
\begin{equation}\label{eq:gammablock}
\gamma =
\begin{pmatrix} n_a & k_q \\ k_q & n_b \end{pmatrix} \oplus
\begin{pmatrix} n_a & k_p \\ k_p & n_b \end{pmatrix} \equiv C_q \oplus C_p
\end{equation}
with $k_q\geq|k_p|$ by local unitary Gaussian operations. The block
structure corresponds to a direct sum of position and momentum space,
i.e., we have reordered $R=(Q_1,Q_2,P_1,P_2)$. Since the normal form in
Eq.(\ref{eq:gammablock}) is unique the parameters $(n_a,n_b,k_q,k_p)$
provide a complete set of local invariants.

The first step towards calculating $E_G$ for these states is to
show that there is always a pure state $\gamma_p$, which is
optimal for Eq.~(\ref{theoEoFG}) and has the same block structure
as $\gamma$ in Eq.~(\ref{eq:gammablock}). To this end we will
first provide a general parameterization for pure state CMs, which
accounts for the direct sum with respect to configuration and
momentum space:

\begin{lemma}
A real symmetric matrix $\gamma_p$ is the covariance matrix of a
pure Gaussian state of $n$ modes iff there exist real symmetric
$n\times n$ matrices $X$ and $Y$ with $X>0$ such that
\begin{equation}\label{GXY}
\gamma_p=\left(\begin{array}{cc}
  X & XY \\
  YX\  & YXY+X^{-1}
\end{array}\right)\;,
\end{equation}
where the block structure corresponds to a direct sum of
configuration and momentum space.
\end{lemma}
 {\it Proof: }A covariance matrix
$\gamma_p$ corresponds to a pure Gaussian state iff
$(\gamma_p\sigma)^2=-\id$. If we write
$$\gamma_p=\left(\begin{array}{cc}
  X & C \\
  C^T  & D
\end{array}\right)$$ with $X,D >0$, then this is equivalent to \begin{eqnarray}
XD&=& \id+C^2,\label{pure1} \\
DC&=&(DC)^T,\label{pure2}\\
CX&=&(CX)^T.\label{pure3}
\end{eqnarray}
Eq.~(\ref{pure3}) implies that $Y:=X^{-1}C
\stackrel{(\ref{pure3})}{=}(X^{-1}C)^T=Y^T$ is indeed symmetric.
Moreover, Eq.~(\ref{pure1}) leads to
$D=X^{-1}(\id+C^2)=X^{-1}+YXY$. Hence, every covariance matrix of
a  pure Gaussian state is of the form in Eq.~(\ref{GXY}).

 Conversely, every such matrix $\gamma_p$ with
$X>0$ is positive definite and has symplectic eigenvalues equal to
one since the spectrum of $-(\gamma_p\sigma)^2$ is the symplectic
spectrum squared of $\gamma_p$. Thus every matrix $\gamma_p$ is an
admissible covariance matrix corresponding to a pure Gaussian
state.\qed

\begin{figure}\epsfig{file=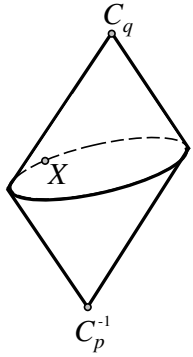,width=2.5cm}
\caption{\label{figcones}For any two-mode Gaussian state with CM
$\gamma=C_q\oplus C_p$ the Gaussian Entanglement of Formation is
given by the entanglement of the least entangled pure state with
CM $\gamma_p=X\oplus X^{-1}$ which is such that $C_p^{-1}\leq
X\leq C_q$. Moreover, the optimal $X$ can be shown to lie on the
rim of the intersection of the forward and backward cones of
$C_p^{-1}$ and $C_q$ respectively.}
\end{figure}

The covariance matrix in the normal form of
Eq.~(\ref{eq:gammablock}) only contains terms which are quadratic
in the momenta but has no linear contributions. This implies that
the state remains invariant under momentum reversal $P\mapsto -P$
and since this can be interpreted as complex conjugation, the
respective density operator is real (in position representation).

Eq.~(\ref{GXY}) gives the covariance matrix of a pure Gaussian
state with respective wave function
\begin{equation}\label{psiXY}\Psi(x)= |\pi X|^{-\frac14}
\exp{\big[-\frac12 x^T(X^{-1}-iY) x \big]},\end{equation} which in
turn becomes real if $Y=0$. The following Lemma shows that for
two-mode states we can, in fact, restrict to these real pure
states in the calculation of $E_G(\gamma)$:

\begin{lemma}\label{blockstructure}
Let $\gamma=C_q\oplus C_p$ be the covariance matrix of a two-mode
 Gaussian state. Then there exists a pure state with covariance matrix
 of the same block structure which minimizes $E_G$ for
 $\gamma$.
\end{lemma}
{\it Proof: } We will show that for every
$\gamma_p(X,Y)\leq\gamma$ of the form in Eq.~(\ref{GXY}) the
covariance matrix $\gamma_p(X,0)$ leads to an improvement for
$E_G$. First note that the block structure of $\gamma$ implies
that $\gamma\geq \gamma_p(X,\pm Y)$ and thus
\begin{eqnarray}
\gamma &\geq&
\frac12\big[\gamma_p(X,Y)+\gamma_p(X,-Y)\big]\\
&=&\gamma_p(X,0)+ 0\oplus(YXY)\\
&\geq& \gamma_p(X,0).
\end{eqnarray}Therefore $\gamma_p(X,0)$ is an admissible
covariance matrix for the $E_G$ optimization problem.

In order to show that $\gamma_p(X,0)$ is less entangled than
$\gamma_p(X,Y)$ we make explicit use of the assumption that we
deal with two-mode states. In this case the entanglement is a
monotonous function of the determinant of the reduced covariance
matrix. The difference of the respective determinants can be
calculated straight forward and it is given by
\begin{eqnarray}\label{detdif}
&&\det\big[\gamma_p^{(A)}(X,Y)\big] -
\det\big[\gamma_p^{(A)}(X,0)\big]\\
 &&=\ Y_{12}^2\ \det\big[ X\big]\ \geq\ 0 ,
\end{eqnarray} which completes the proof.\qed

According to Lemma \ref{blockstructure}  the remaining task for
calculating $E_G$ is to find the CM $\gamma_p=X\oplus X^{-1}$
which has minimal entanglement under the constraint that
\begin{equation}\label{constraint1}C_p^{-1} \le X \le
C_q.\end{equation} This inequality has a simple graphical
depiction stemming from the fact that the set of positive
semi-definite matrices $X$ satisfying an inequality as,
e.g. $C_q-X\geq 0$, form a cone,
which is equivalent to the
(backward) 
light cone
of $C_q$ 
in Minkowski space: if we expand a
Hermitian $2\times2$ matrix in terms of Pauli matrices (and the
identity), the expansion coefficients play the role of the
space-time coordinates and the Minkowski norm is simply given by
the determinant of the matrix. Hence, by Eq.(\ref{constraint1})
$X$ has to lie in the backward cone of $C_q$ and in the forward
cone of $C_p^{-1}$ (see Fig.\ref{figcones}).

Instead of minimizing the entropy of the reduced state under this
constraint, we may as well minimize the determinant of one of the
local covariance matrices
\begin{eqnarray}\label{mX}
m(X) &\equiv& 1 + \frac{X_{12}^2}{\det X} = 1 +
\frac{(X^{-1})_{12}^2}{\det (X^{-1})}\\
&=& X_{11}(X^{-1})_{11}\ ,
\end{eqnarray}
since, as already stated, this is a monotonously increasing
function of the entanglement. Thus we have to find
\begin{equation}\label{eq:generalmin}
\min_X \left\{ m(X) \bigm| C_p^{-1}\le X \le C_q \right\}
\end{equation}
over the real, symmetric 2$\times$2 matrices $X$.

In fact, for the optimal $X$ both inequalities have to be
saturated, i.\,e.\
\begin{equation}\label{detzero}
\det{(C_q-X)}=\det{(X-C_p^{-1})}=0.
\end{equation} In order to see this assume we are given a
matrix $X$ with $C_p^{-1}\le X<C_q$. Then we can decrease the
value of $m(X)$ with a matrix $\hat{X}:=X+\epsilon \id$ by
increasing $\epsilon>0$ until $C_q-\hat{X}$ is of rank 1. However,
by Eq.~(\ref{mX}) the same argument holds for $X^{-1}<C_p$.

To depict it geometrically again, the optimal $X$ has to lie on
the rim given by the intersection of the backward and forward
cones of $C_q$ and $C_p^{-1}$ respectively. Hence, we have reduced
the number of free parameters in the calculation of $E_G(\gamma)$
to one angle, which parameterizes the ellipse of this
intersection.

For every explicitly given CM $\gamma$ minimizing $m(X)$ on this
ellipse is now straight forward. Writing down the resulting value
for $E_G$ in terms of the general parameters $(n_a,n_b,k_q,k_p)$
of Eq.~(\ref{eq:gammablock}) leads, however, to quite cumbersome
formulae involving the roots of a forth order polynomial. Since
not much insight is coming out of these expressions we refrain
from writing them down explicitly and continue with discussing the
special case $n_a=n_b$ for which we obtain a simple formula for
$E_G$.

Nevertheless, for an arbitrary but explicitly given $\gamma$ the
remaining variation under the constraint in Eq.(\ref{detzero}) is
a simple exercise which can be solved analytically with the help
of any computer algebra program. For some examples, see
Sec.~\ref{expSec}.

\section{Symmetric states}\label{symSec}

Symmetric two-mode Gaussian states with CM of the form in
Eq.~(\ref{eq:gammablock}) with $n_a=n_b\equiv n$ arise naturally
when the two beams of a two-mode squeezed state are sent through
identical lossy fibers \cite{KW87} (see also Sec.\ref{expSec}).
The entanglement of formation $E_F$ of these states was calculated
in \cite{GWKWC03} and it was proven that a decomposition into
Gaussian states gives rise to the optimal value. Together with the
obvious fact that $E_G$ is an upper bound for $E_F$ this implies
that $E_G=E_F$ in this case. Since the calculation of $E_F$ is
however quite technical and in order to make the present article
more self-contained, we provide in the following a simpler way to
obtain $E_G$.

In principle we could utilize the general results of the
previous section, which simplify greatly for the symmetric
case. However, we give an alternative proof and
reduce the result to the fact that the optimal $\gamma_p$ in
Eq.~(\ref{theoEoFG}) has the same {\it logarithmic negativity}
\cite{VW02} as $\gamma$. A similar argument is used in
Sec.~\ref{addSec} to prove additivity of $E_G$.
\begin{prop}[$E_G$ for symmetric states]\label{eofsym}
For symmetric $1\times1$ Gaussian states, i.e. states whose CM
$\gamma$ is characterized by local invariants
$(n,n,k_q,k_p)$, the Gaussian entanglement of
formation is given by
\begin{equation}
\Eg(\gamma) = H(r_0),
\end{equation}
where the minimum two-mode squeezing required is given by
\begin{equation}
r_0 = \frac{1}{2}\ln[(n-k_q)(n+k_p)]
\end{equation}
and $H(r)$ is defined in Eq.~(\ref{Hr}).
\end{prop}
{\it Proof: } First, instead of $\gamma$ we consider the locally
equivalent CM $\gamma'$ which is obtained from $\gamma$ by
squeezing \footnote{``Squeezing by $\lambda$'' describes the local
  unitary operation that (in the Heisenberg picture) multiplies
  (divides) the operators $Q_{A,B}$ ($P_{A,B}$) by $\lambda$.}
both $Q_A$ and $Q_B$ by $\lambda = [(n+k_p)/(n-k_q)]^{1/4}$.
Clearly the CM $\gamma'$ has the same $\Eg$ as $\gamma$. It is
straightforward to check that the pure two-mode squeezed state
with two-mode squeezing parameter $r_0$ and corresponding CM
$\gamma(r_0)$ is indeed smaller than $\gamma'$.

That there can be no pure state $\gamma_p$ with less entanglement
satisfying $\gamma_p\leq\gamma'$ follows from the monotonic dependence
of pure state entanglement on the two-mode squeezing parameter: any
pure two-mode Gaussian state is locally equivalent to a two-mode
squeezed state $\gamma(r)$ and its entanglement is given
by $H(r)$.
An important entanglement-related characteristic of these CMs are the
\emph{symplectic eigenvalues} of the partially transposed CM
\cite{VW02}, in particular those smaller than one. They are invariant
under local unitary Gaussian operations and for the two-mode squeezed state
given by $e^{\pm r}$. For the symmetric CM $\gamma$ they are
$\sqrt{(n\mp k_q)(n\pm k_p)}$. Thus the smallest symplectic eigenvalues
of $\gamma$ and $\gamma(r_0)$ coincide.

For positive matrices $A\geq B$ implies $a_k\geq b_k$, where $a_k$
($b_k$) denote the ordered symplectic eigenvalues of $A$ ($B$)
\cite{GECP03}. Since the ordering $A\geq B$ is preserved under partial
transposition, all pure states with less entanglement than
$\gamma(r_0)$ cannot possibly satisfy $\gamma\geq\gamma_p$, hence
$r_0$ is optimal.\qed

Thus for symmetric states $\gamma$ the optimal pure state
$\gamma_p$ is characterized by the fact that the smallest
symplectic eigenvalues $s_1(\tilde\gamma)$ and
$s_1(\tilde\gamma_p)$ of the two partially transposed CMs are
identical. According to \cite{VW02} this implies that the
logarithmic negativity,
$E_N(\gamma)=-\frac{1}{2}\ln[s_1(\tilde\gamma)]$ of both states is
the same, i.e., in the optimal decomposition pure Gaussian states
are mixed such that ``no negativity is lost'' in the mixing
process. For non-symmetric states this is no longer possible and
$s_1(\tilde\gamma)$ is strictly larger than $s_1(\tilde\gamma_p)$,
i.e. more entanglement is needed to form $\gamma$ than required by
its negativity.

\section{Additivity}\label{addSec}
 One longstanding question about the entanglement of
formation is if it is \emph{additive}, that is whether
$E_F(\rho_1\otimes\rho_2)=E_F(\rho_1)+E_F(\rho_2)$ or whether one
may get an ``entanglement discount'' when generating several
states at a time. Here we show that for \emph{symmetric}
 Gaussian states the \emph{Gaussian} entanglement of
formation $E_G$ is additive. Since for these
states $E_G$ was shown \cite{GWKWC03} to equal $E_F$ this may hint
at additivity of even the latter quantity for Gaussian states.

\begin{prop}[$\Eg$ is additive for symmetric states]\label{propsym}
Let $\gamma_l, l=1,\dots, N$ describe symmetric Gaussian states with
local invariants $(n_l,n_l,k_{q,l},k_{p,l})$ and let
$\Gamma=\oplus_{l=1}^N\gamma_l$ describe the tensor product of these
states, then
\begin{equation}
\Eg(\Gamma) = \sum_l\Eg(\gamma_l).
\end{equation}
\end{prop}
{\it Proof: }
Let the logarithmic negativity of the $k$th state be $r_k$,
and assume $r_{k+1}\leq r_k$. To show additivity, we use again
that $A\geq B>0$ implies $a_k\geq b_k$ for the ordered symplectic
eigenvalues of $A(B)$.

Let $\Gamma_p\leq\Gamma$ be a pure $N\times N$-mode CM. Consider
the partially transposed CM  $\tilde\Gamma$. We have
\[\tilde\Gamma\geq\tilde\Gamma_p,\]
which implies $s_k\geq s_k'$, where $\{s_k,k=1,...,N\}$ denote the
(descendingly ordered) symplectic eigenvalues of $\tilde\Gamma$,
and $s_k'$ the same for $\tilde\Gamma_p$.

All pure bipartite Gaussian states are locally equivalent to a
tensor product of two-mode squeezed states \cite{HW01} with
(ordered) two-mode squeezing parameters $t_k$. For the following
we only need to look at the $N$ smallest symplectic eigenvalues.
For these we have
\[s'_k=e^{-t_k}\,\,\,\mbox{for $\Gamma_p$}
\]
and
\[
s_k = e^{-r_k}\,\,\, \mbox{for $\Gamma$}.
\]
Hence $\Gamma\geq\Gamma_p$ implies  $t_k\geq r_k$, i.e., the
optimal joint decomposition is the tensor product of the optimal
decompositions for the individual copies. Thus $E_G$ is additive.\qed

For the non-symmetric case the optimal individual decomposition
does no longer allow $t_k=r_k$, i.e. more entanglement than
required by the logarithmic negativity must be expended to produce
$\rho$. Therefore, the previous argument does not hold and the
question of additivity of $E_G$ remains open for general
$1\times1$ Gaussian states.

\section{Examples}\label{expSec}

In this section we will apply the Gaussian Entanglement of
Formation to a simple practical example. Consider a two-mode
squeezed state (TMSS) with CM
\begin{equation}\label{eq:cmsympara1}\begin{gathered}
\gamma = \begin{pmatrix}  c & 0 & - s & 0 \\ 0 &  c & 0 &  s
\\ - s & 0 &  c & 0 \\ 0 &  s & 0 &  c
\end{pmatrix}, \\
c = \cosh{2 r},\quad s = \sinh{2 r},
\end{gathered}\end{equation}
which is to be distributed between two parties by means of a lossy
optical fiber. There are two extremal settings for the
transmission of the state which may lead to different values for
the distributed entanglement: The source could be placed either at
one party's site (``asymmetric setting'') or halfway between both
parties (``symmetric setting''). In the former case, one mode is
transmitted through the whole length of the fiber while the other
one is retained unaffected, in the latter both modes are
transmitted through half the length of the fiber each. It turns
out that depending on thermal noise and transmission length, one
or the other setting yields more entanglement for the distributed
state from a given squeezing of the initial state.

According to
\cite{DG97}
the Gaussian state after transmission through the fiber has a CM
\begin{gather}\label{eq:cmfibpara}
\gamma' = \begin{pmatrix} c_1' & 0 & -s' & 0 \\ 0 &  c_1' & 0 &
s'
\\ -s' & 0 &  c_2' & 0 \\ 0 &  s' & 0 &  c_2'
\end{pmatrix} \\
\begin{aligned}
& \text{ with } c_i' = c\, T_i^2 + (2 N_\text{th}+ 1)\, (1-T_i^2) \text{ for } i=1,2 \\
& \text{ and } s' = s\, T_1 \, T_2.
\end{aligned} \notag
\end{gather}
Here $T_1$, $T_2$ are the transmission coefficients of the fiber
for the respective modes and the fiber is in a thermal state with
mean number of photons $N_\text{th}$, which is in turn related to
a ``temperature'' $\tau$ by  $N_\text{th}=(\exp(1/\tau)-1)^{-1}$.
Note that in quantum optical settings using optical frequencies we
have $\tau=0$.

\begin{figure}[t]
\psfrag{E}{$E_G(\rho)$}
\psfrag{2}2\psfrag{4}4\psfrag{6}6\psfrag{8}8\psfrag{10}{10}\psfrag{0.1}{0.1}\psfrag{0.2}{0.2}\psfrag{0.3}{0.3}\psfrag{0.4}{0.4}\psfrag{0.5}{0.5}\psfrag{1.5}{1.5}\psfrag{2.5}{2.5}
\psfrag{l}[b][B]{\raisebox{1.2ex}{\hspace*{1em}$l/l_\text{A}$}}
\epsffile{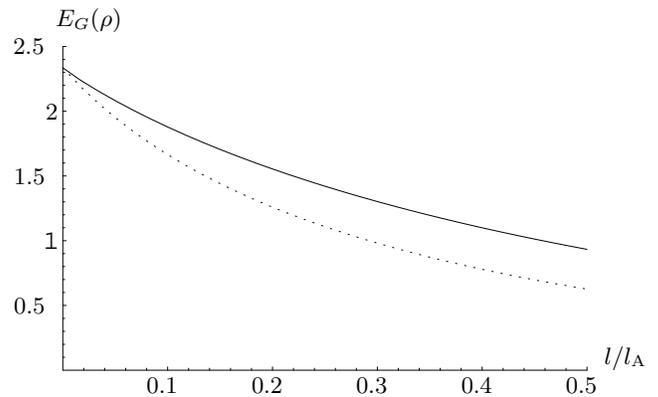} \caption{Gaussian Entanglement of
Formation (in units of ebits) for a TMSS with $r=1$ after
transmission through a lossy optical fiber at temperature
$\tau=0$. The plot shows $E_G(\rho)$ versus transmission length
$l/l_\text{A}$ for the symmetric and asymmetric setting (dotted
and solid line, respectively).} \label{fig:fiberzero}
\end{figure}

\begin{figure}[t]
\psfrag{E}{$E_G(\rho)$}\psfrag{1}1\psfrag{2}2\psfrag{3}3\psfrag{4}4\psfrag{0}{0}\psfrag{0.1}{0.1}\psfrag{0.2}{0.2}\psfrag{0.3}{0.3}\psfrag{0.4}{0.4}\psfrag{0.5}{0.5}\psfrag{1.5}{1.5}\psfrag{2.5}{2.5}
\psfrag{l}[b][B]{\raisebox{1.2ex}{\hspace*{1em}$l/l_\text{A}$}}
\epsffile{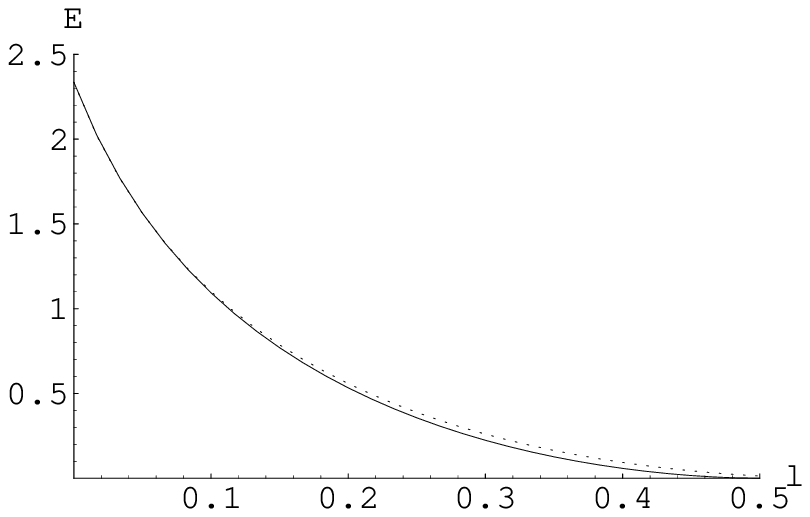} \caption{Gaussian Entanglement of
Formation  (in units of ebits) for a TMSS with $r=1$ after
transmission through a lossy optical fiber at temperature
$\tau=1$. The plot shows $E_G(\rho)$ versus transmission length
$l/l_\text{A}$ for the symmetric and asymmetric setting (dotted
and solid line, respectively).} \label{fig:fiberfinite}
\end{figure}

 Depending
on the setting, the transmission coefficients take on the values
$T_1=e^{-l/l_\text{A}}, T_2=1$ (asymmetric setting) or
$T_1=T_2=e^{-l/(2 l_\text{A})}$ (symmetric setting) where
$l/l_\text{A}$ denotes the total length $l$ of the fiber in units
of the absorption length $l_\text{A}$. As an example, we compare
the two settings for a TMSS with $r=1$. While for temperature
$\tau=0$ (Fig.~\ref{fig:fiberzero}) the asymmetric setting always
yields a higher entanglement for the final state, at finite
temperature $\tau=1$ (Fig.~\ref{fig:fiberfinite}) the symmetric
setting is to be preferred for longer ranges.

\begin{figure}[t]
\psfrag{e}{$E_G$}\psfrag{1}1\psfrag{0.5}{0.5}\psfrag{1.5}{1.5}\psfrag{2.5}{2.5}\psfrag{2}2\psfrag{3}3\psfrag{4}4\psfrag{0}0\psfrag{5}5\psfrag{5}5\psfrag{10}{10}\psfrag{r}{r}\psfrag{0.1}{0.1}\psfrag{0.2}{0.2}
\psfrag{l}[b][B]{\raisebox{1.2ex}{\hspace*{1em}$l/l_\text{A}$}}
\epsfxsize=8.5cm\epsffile{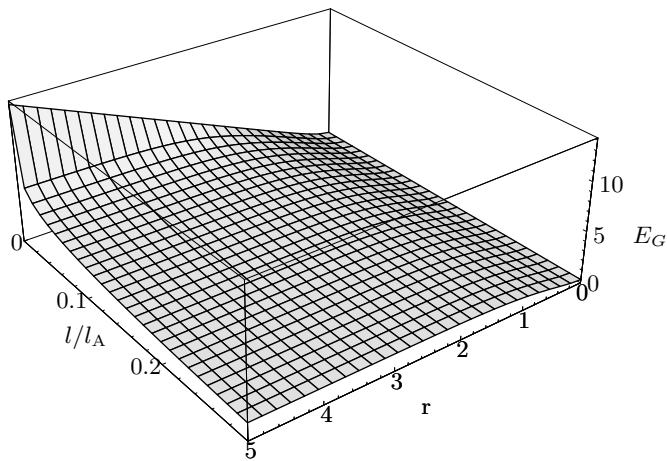} \caption{Gaussian
Entanglement of Formation (in units of ebits) for a TMSS with
initial squeezing $r$ after symmetric transmission through a lossy
optical fiber at zero temperature.  $l/l_\text{A}$ is the
transmission length in units of the absorption length.}
\label{fig:fiber3d}
\end{figure}

Fig.~\ref{fig:fiber3d} shows $E_G$ as a function of the initial
squeezing $r$ and the transmission length $l$ for the symmetric
setting at zero temperature. As already indicated in
\cite{SOW00,SW01} increasing the squeezing over a certain
threshold has only a negligible effect on the transmitted
entanglement already after a small fraction of the absorption
length.

\section{Conclusion}\label{conSec}

We introduced a Gaussian version of the Entanglement of Formation
by taking into account only decompositions into pure Gaussian
states. On Gaussian states this is a proper entanglement measure
in the sense that it is non-increasing under GLOCC operations.
Moreover, it is an upper bound for the full Entanglement of
Formation which is tight and additive at least for symmetric
two-mode Gaussian states. However, it remains an open question
whether this is true for all Gaussian states.\vspace*{10pt}

We have shown how to analytically calculate $E_G$ for all two-mode
Gaussian states and given a simple formula for the symmetric case.
For multimode bipartite states,
the (numerical) computation of $E_G$ becomes rapidly more difficult:
Although Proposition 1 still allows to
replace the minimization over all possible decomposition into pure
Gaussian states by the minimization over the (for the $n\times n$
case) $2n(2n+1)$ parameter set of pure $2n$-mode covariance matrices
the further simplifications used in the $1\times1$ case are no longer
available and the minimization is no longer easy. To compute $E_G$ for
interesting multi-mode states such as two copies on non-symmetric
$1\times1$ states or bound entangled states \cite{WW01a} new methods
need to be developed.

GG and MMW thank Pranaw Rungta for interesting discussion during the
ESF QIT conference in Gdansk, 2001. Funding by the European Union
project QUPRODIS and the German National Academic Foundation (OK) and
financial support by the A2 and A8 consortia is gratefully
acknowledged.\vspace*{10pt}

\appendix

\section{Proof of Lemma 3}\label{app1Sec}

The following Lemma about decompositions of classical multivariate
Gaussian distributions is used in the proof of Prop.
\ref{propgen}.
\begin{lemma}\label{classGauss}
Let
\begin{equation}\label{GAax}
  G(A,a,x):=|A|^{\frac12} (2\pi)^{-\frac{n}2}\ e^{-\frac12
 (x-a)^T A(x-a)},\quad x,a\in\mathbb{R}^n
\end{equation}
be a Gaussian probability distribution with symmetric $A > 0$ and
consider an arbitrary convex decomposition of $G(A,0,x)$ into
other Gaussian distributions of this form:
\begin{equation}\label{Gdecomp}
  G(A,0,x)=\int\mu(dB,db)G(B,b,x).
\end{equation}
Then all the distributions contributing to this decomposition have
to satisfy $B\geq A$ in the sense that
\begin{equation}\label{measzero}
\mu\Big(\big\{(B,b)\ \big|\ \exists_x:
x^T(A-B)x>0\big\}\Big)=0.
\end{equation}
\end{lemma}

{\it Proof: } Let us first define a set
\begin{eqnarray}\label{S1}
S(x,\epsilon)&:=&\Big\{(B,b)\ \big|\  x^TAx -
\epsilon ||x||^2 \geq  x^TB x \Big\}\label{S2}.
\end{eqnarray}
Integrating Eq.~(\ref{Gdecomp}) only over $S(x,\epsilon)$ leads
then to a lower bound on $G(A,0,x)$:
\begin{equation}\label{ineq1}
G(A,0,x)\geq\int_{S(x,\epsilon)}\hspace{-10pt}\mu(dB,db)G(B,b,x).
\end{equation}
Inserting $G(A,0,x)$ and $G(B,b,x)$ we can symmetrize inequality
(\ref{ineq1}) with respect to $x\mapsto-x$, which leads to
\begin{eqnarray}\label{Eq137}
|A|^{\frac12} e^{-\frac12  x^TAx}&\geq& \int_{S(x,\epsilon)}\hspace{-10pt}\mu(dB,db)\ |B|^{\frac12} \cosh{( x^TBb)}\nonumber \\
&& \quad e^{-\frac12( x^TBx + b^TBb)} .
\end{eqnarray}
Utilizing $\cosh\geq 1$ and the defining property of the set
$S(x,\epsilon)$ we have
\begin{equation}\label{finalineq}
|A|^{\frac12} e^{-\frac{\epsilon}2
||x||^2}\geq\int_{S(x,\epsilon)}\hspace{-10pt}\mu(dB,db)\
|B|^{\frac12} e^{-\frac12  b^TBb}.
\end{equation}
Note that the r.h.s. of Eq.~(\ref{finalineq}) does no longer depend
on the norm of $x$ but merely on its angular components. Taking
the limit $||x||\rightarrow\infty$ implies then that
$S(x,\epsilon)$ is of measure zero for every $\epsilon>0$.
Moreover, every countable union of such sets is of measure zero.
In particular:
\begin{eqnarray}
&& \bigcup_{x\in\mathbb{Q}^n} \bigcup_{m\in\mathbb{N}}
S(x,\frac1m)\ =\\
&=& \Big\{(B,b)\big|\exists x\in\mathbb{Q}^n, \exists
m\in\mathbb{N}:\nonumber\\ && \qquad x^TAx\geq
x^TBx+\frac1m ||x||^2 \Big\}\\
&=& \Big\{ (B,b)\big| \exists x\in \mathbb{Q}^n :
x^TAx > x^TBx \Big\}\\
&=& \Big\{(B,b)\big| \exists x\in\mathbb{R}^n : x^T(A-B)
x > 0\Big\},
\end{eqnarray} where we have of course used that $\mathbb{Q}^n$ is dense in
$\mathbb{R}^n$.\qed

\end{document}